\documentclass[twocolumn,showpacs,preprintnumbers,amsmath,amssymb]{revtex4}
\usepackage{graphicx}
\usepackage{multirow}
\usepackage{color}

\newcommand{\ket}[1]{\ensuremath{\left|#1\right>}}

\begin{document}

\title{Internal-state thermometry by depletion spectroscopy in a cold guided beam of formaldehyde}

\author{M. Motsch}
\author{M. Schenk}
\author{L.D. van Buuren}
\author{M. Zeppenfeld}
\author{P.W.H. Pinkse}
\author{G. Rempe}
\affiliation{Max-Planck-Institut f{\"u}r Quantenoptik, Hans-Kopfermann-Str. 1, 85748 Garching, Germany}

\begin{abstract}
We present measurements of the internal state distribution of electrostatically guided formaldehyde. Upon excitation with continuous tunable ultraviolet laser light the molecules dissociate, leading to a decrease in the molecular flux. The population of individual guided states is measured by addressing transitions originating from them. The measured populations of selected states show good agreement with theoretical calculations for different temperatures of the molecule source. The purity of the guided beam as deduced from the entropy of the guided sample using a source temperature of 150\,K corresponds to that of a thermal ensemble with a temperature of about 30\,K.
\end{abstract}


\maketitle

Cold polar molecules open new roads for studies of dipolar collisions, cold chemistry \cite{Balakrishnan:ColdChem,Krems:ColdChem} and quantum computation \cite{DeMille:QuantumComp}. Cold molecules are also expected to increase the sensitivity in precision measurements as in e.g. tests of parity violation \cite{Daussy:Parity} or the search for the electron dipole moment \cite{Hudson:EDM,DeMille:EDM}. Spectroscopy in cold and slow beams offers the advantage of long interaction times which increases the achievable resolution \cite{Hudson,vanVeldhoven}. So far a variety of methods have been developed which typically reach temperatures around 1\,K for naturally occurring molecules \cite{Doyle:QuoVadis}. To reach lower temperatures, thereby bridging the gap to the ultracold regime, new trapping and cooling schemes are necessary. Suggested schemes include sympathetic cooling with cold alkali atoms \cite{Schlunk:Rb,Rieger:Rb} and cavity cooling \cite{Chan:CavCoolExp,Maunz:CavCoolExp,Nussmann:CavCoolExp} of external \cite{Horak:CavCoolProposal,Vuletic:CavCoolProposal,Lev:CavCoolProposal,Lu:CavCoolProposal} and internal \cite{Morigi} degrees of freedom. Detailed knowledge of the internal state distribution of trapped or guided molecules is essential for the implementation of these schemes as well as for investigation of collisions and ultracold chemistry \cite{Avdeenkov}.

One approach for producing high fluxes of cold polar molecules is velocity filtering. Continuous fluxes of up to $10^{10}\,\mathrm{s^{-1}}$ have been demonstrated e.g. for formaldehyde (H$_2$CO) and deuterated ammonia (ND$_3$). The molecules have longitudinal velocities of 0--100\,m/s, corresponding to a translational temperature of $\approx$\,5\,K \cite{Rangwala,JunglenEPJD}. So far the internal state distribution was not accessible in the experiments, since only the signal integrated over all states was measured.

Here we present a new method which allows to measure the contribution of individual states in a cold guided beam of H$_2$CO even when only the integrated signal is accessible to the detector. The method is therefore well suited to be combined with state integrating devices like electron ionization or Langmuir-Taylor detectors. By depletion spectroscopy we probe the internal state distribution of formaldehyde molecules guided in an electrostatic quadrupole field. The molecules are excited in the $2^1_04^3_0$ vibrational band of the $\tilde X^1A_1\to\tilde A^1A_2$ transition around 330\,nm \cite{Clouthier}. Since the $\tilde A^1A_2$ excited state of formaldehyde predissociates \cite{Moore}, molecules pumped to this state are lost from the guided beam. Hence, the depletion induced by the laser beam is a direct measure for the population of a state.

The technique is not restricted to formaldehyde. It is not even necessary for the molecular excited state to be coupled to a dissociative channel. Upon excitation, the molecule will decay back to a different ground state with a likely smaller Stark shift or even to a high-field-seeking state and be lost from the guide, also causing a depletion signal. Compared to state-selective ionization techniques like resonance-enhanced multiphoton ionization REMPI, our method has the advantage of smaller complexity and of being well adapted to continuous beams.

Velocity filtering is based on selection of the slowest molecules from an effusive source as previously described in detail \cite{JunglenEPJD}. Transversely slow molecules in low-field-seeking states are trapped in two dimensions by an electrostatic guide. Selection of longitudinal velocity is obtained by bending the guide. For each rotational state a longitudinal and transverse cut-off velocity exists depending on its Stark shift, the maximum trapping field in the quadrupole guide and the bend radius. The guided signal can be increased by cooling the effusive source. Cooling reduces the mean thermal velocity of the molecules in the source, which increases the fraction of molecules below the cut-off velocity for a specific rotational state. Furthermore the internal state distribution of the molecules becomes narrower, leading to a purer beam.

\begin{figure}
\begin{centering}
\includegraphics[scale=1.]{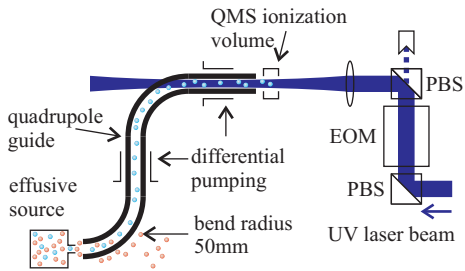}
\caption{(color online) Experimental setup. Formaldehyde molecules from an effusive source are injected into the quadrupole guide and led to the QMS cross-beam ionization volume. In the last 15\,cm long straight section the UV beam is overlapped with the molecular beam. The ligth is switched on and off by an EOM placed between two polarizers.}
\label{fig:setup}
\end{centering}
\end{figure}

The experimental setup is shown in Fig.\;\ref{fig:setup}. Formaldehyde gas is injected into the quadrupole guide through a 1.5\,mm diameter ceramic nozzle. The temperature of the liquid nitrogen cooled nozzle can be varied in the range 100--300\,K by a heater. The molecules are guided around two 50\,mm radius bends and through two differential pumping stages to the cross-beam ionization unit of a quadrupole mass spectrometer (QMS) placed in an ultrahigh vacuum chamber with a background pressure in the $10^{-11}$\,mbar regime for detection. In the straight section before the QMS ionization unit an ultraviolet (UV) laser beam is overlapped with the guided molecular beam over a length of $\approx$\,15\,cm.

The narrow-band tunable UV laser light is created by second-harmonic generation of light from a ring dye laser in an external enhancement cavity. The dye laser is locked to a temperature-stabilized reference cavity and its output frequency is monitored by a wavemeter. After beam shaping and laser power switching by an electro-optical modulator, around 100\,mW  laser light at a wavelength of 330\,nm is available for the experiment. The laser beam is focussed into the guide counter-propagating the molecules. The beam waist of 150--200\,$\mu$m is optimized for minimum UV light scattering on the high voltage electrodes, which would lead to a pressure rise and consequently an increase in background count rate. After careful alignment no significant influence of the UV light on the QMS count rate is observed. Numerical simulations of the motion of molecules in the trapping field show a typical diameter of the guided beam of $\approx$\,0.8\,mm, independent of the rotational state, causing most molecules to pass through the laser beam on their orbit.

The internal state distribution of guided molecules is determined by the thermal population in the effusive source including nuclear spin statistics and by the Stark shift of the rotational states in the applied guiding fields. The Stark shift $\Delta W_s$ is calculated by numerical diagonalization of the asymmetric rotor Hamiltonian in the presence of an electric field \cite{Hain}. The signal contribution of a rotational state \ket{J,\tau,M} is proportional to the square of its Stark energy $(\Delta W_s(|\vec{E}|))^2$ at the maximum trapping field \cite{JunglenEPJD}. $J$ is the rotational quantum number, $\tau$ is a pseudo quantum number labeling states, and $M$ is the projection of the angular momentum on the electric field vector. Stark shifts at a typical trapping field of 100\,kV/cm and populations of rotational states in the guide are listed in Tab.\;\ref{tab:pop} for states which contribute more than 3\% to the total flux at a source temperature of 150\,K.

\begin{table}
\begin{centering}
\begin{tabular}{|c|c|c|c|}
\hline
Rotational & $\Delta W_S$ & Source & Guided\\
State & $\mathrm{[cm^{-1}]}$ & Pop. [\%] & Pop. [\%] \\
\ket{J,\tau,M} & $@100\,\mathrm{kV/cm}$ & $@150\,\mathrm{K}$ & $@100\,\mathrm{kV/cm}$ \\
\hline
$\ket{3,3,3}$ & 2.86 & 0.252 &  8.52 \\
$\ket{4,2,4}$ & 2.28 & 0.230 &  4.93 \\
$\ket{1,1,1}$ & 1.29 & 0.531 &  3.62 \\
$\ket{3,3,2}$ & 1.82 & 0.252 &  3.46 \\
$\ket{2,2,2}$ & 2.42 & 0.134 &  3.22 \\
$\ket{5,1,5}$ & 1.90 & 0.205 &  3.05 \\
\hline
\end{tabular}
\caption{Calculated Stark shifts $\Delta$W$_s$ and population of selected rotational states \ket{J,\tau,M} in the source and in the guide. Note the increase of population of these states relative to their contribution in the source by a factor of 10 to 35.}
\label{tab:pop}
\end{centering}
\end{table}

Room-temperature absorption spectroscopy has been performed for identification of transitions from different rotational states \ket{J,\tau} of formaldehyde \cite{Motsch}. The rotational constants derived from a fit to this data using genetic algorithms \cite{Meerts:GeneticAlgorithm} are used for line identification and prediction of line positions for guided states.

\begin{figure}
\begin{centering}
\includegraphics[scale=1]{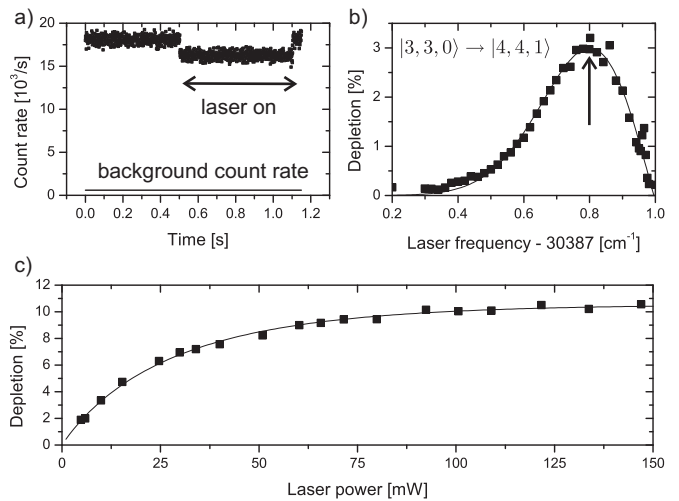}
\caption{(a) Time-resolved ion count rate measured with the QMS. The UV light is on between 0.5 and 1.1\,s. (b) The transition $\ket{3,3,0}\rightarrow\ket{4,4,1}$ measured at an electrode voltage of $\pm5\,\mathrm{kV}$. The solid line is a fit to the measured line shape taking into account the laser beam and molecular beam profile and the electric field distribution in the spectroscopy section. (c) Laser power dependence of the depletion signal. The laser frequency is fixed to the maximum of the transition in the guiding field as indicated in (b). The solid line shows the result of a fit using $\Delta(P)$. For (c) the electrode voltage was reduced to $\pm1$\,kV.}
\label{fig:signal}
\end{centering}
\end{figure}

Upon excitation with the UV light the molecules dissociate and a decrease in the QMS signal is detected as shown in Fig.\;\ref{fig:signal}\,(a). By changing the laser frequency, different transitions are addressed. In the experiment we vary the laser power to obtain the saturated depletion signal. Here the laser frequency is fixed to the maximum of the transition in the guiding field (see Figs.\;\ref{fig:signal}\,(b) and (c)). The saturated depletion then allows extraction of the population of the specific rotational state addressed. To reach saturation with the available laser power, the electrode voltage is set to a relatively small value of $\pm1 $\,kV, resulting in smaller line broadenings.

Several effects lead to line broadenings and line shifts of the depletion signal, such that the depletion peaks are not centered at the zero-field transition frequency (see Fig.\;\ref{fig:signal}\,(b)). Since the molecules move with a typical velocity $\approx$\,90\,m/s, they experience on average a Doppler shift of $\approx$\,-300\,MHz. Moreover, the interaction time with the laser beam is velocity dependent, giving rise to a $1/v$ absorption probability for low laser intensities. The laser power dependence of the depletion signal can be described by $\Delta(P)=N_0\cdot[1-\int^{v_{l,max}}_0 n(v)\exp(-\alpha P l/v)dv]$ where $N_0$ represents the relative population of the specific state addressed by the laser, $n(v)$ is the normalized longitudinal velocity distribution, $P$ is the laser power, $\alpha$ is a transition strength factor and $l$ is the interaction length.

The largest effect on the line shape originates from the radial position dependence of the electric field experienced by the molecules and the resulting Stark shifts. Taking into account the radial distribution of the molecular beam derived from simulations and the laser beam diameter, the line shape can be reproduced (see solid line in Fig.\;\ref{fig:signal} (b)).

Since the electric fields in the guide vary in strength and direction, it is not possible to irradiate the molecules with a well-defined polarization with respect to the guiding electric field. If for a given state \ket{J,\tau} more than one $M$ component contributes to the molecular flux,  this leads to additional line broadening and makes it impossible to depopulate the different $M$ states independently. Instead a signal integrated over all guided $M$ substates of the \ket{J,\tau} state is measured.

For depletion spectroscopy we mainly use the R-branch out of the $2_0^14_0^3$ vibrational band since there the frequency difference between transitions is smallest, allowing easier tuning of the laser. The transitions used and their zero-field frequencies are given in Tab.\;\ref{tab:trans}. Surprisingly, we observe a line splitting of the transitions from state \ket{1,1} and \ket{3,3} of $\approx$\,0.04\,cm$^{-1}$ and $\approx$\,0.1\,cm$^{-1}$, respectively, persisting in the low-field limit. The two doublet components show the same saturation limit also for varying source temperature. Therefore they most likely originate from the same guided state. The cause for this splitting might be a coupling to highly excited rovibrational states of the $\tilde X^1A_1$ potential energy surface \cite{Polik}.

\begin{table}
\begin{centering}
\begin{tabular}{|c|c|c|}
\hline
\multirow{2}{*}{\ket{J,\tau}} & $\tilde X^1A_1 \rightarrow \tilde A^1A_2$ & \multirow{2}{*}{Frequency\,[cm$^{-1}$]}\\
& \ket{J,K_{-1},K_{+1}} $\rightarrow$ \ket{J',K'_{-1},K'_{+1}} & \\
\hline
\ket{1,1} & \ket{1,1,0} $\rightarrow$ \ket{2,2,1} & 30364.38\\
\ket{2,2} & \ket{2,2,0} $\rightarrow$ \ket{3,3,1} & 30377.26\\
\ket{3,3} & \ket{3,3,0} $\rightarrow$ \ket{4,4,1} & 30387.98\\
\ket{4,2} & \ket{4,3,1} $\rightarrow$ \ket{5,4,2} & 30388.76\\
\ket{5,1} & \ket{5,3,2} $\rightarrow$ \ket{5,4,1} & 30376.62\\
\ket{5,5} & \ket{5,5,0} $\rightarrow$ \ket{6,6,1} & 30403.84\\
\hline
\end{tabular}
\caption{The transitions used to address the highest populated states in the guide. The states are labeled by $K_{-1}$ and $K_{+1}$, the $K$ quantum numbers of the projection of $J$ on the figure axis in the limiting case of the prolate and oblate symmetric top.}
\label{tab:trans}
\end{centering}
\end{table}

\begin{figure}
\begin{centering}
\includegraphics[scale=1.]{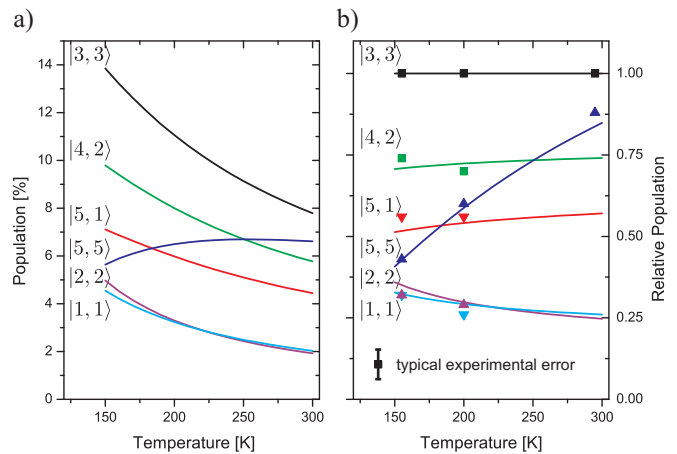}
\caption{(color online) Temperature dependence of population for selected guided states \ket{J,\tau}. (a) Calculated population of states. (b) Experimentally determined and calculated population relative to state \ket{3,3}.}
\label{fig:popT}
\end{centering}
\end{figure}

The relative population of individual guided states, normalized to the state with maximum contribution, \ket{3,3}, is shown in Fig.\;\ref{fig:popT} and is in good agreement with our calculations. However, for $\pm$\,1\,kV electrode voltage, the measured absolute population of the states compared to the calculated population is systematically lower by 20\%, independent of the addressed state. We attribute this difference to molecules spiraling in elliptic orbits which never become resonant with the laser beam, and to molecules which do not dissociate and decay back to a different guided ground state. At the used wavelength of $\approx$\,330\,nm the quantum efficiency for dissociation is $>$\,90\,\% \cite{Moore}. 30--40\,\% of these undissociated molecules could still be guided as estimated from the Stark shifts of rotational states to which they can decay.

By varying the source temperature between 150 and 300\,K, the internal state distribution of the formaldehyde molecules as well as the velocity distribution in the source is modified. For decreasing inlet temperature, the reduction of the molecules' mean velocity is directly evident from an increase in guided flux. Temperature-dependent measurements of the internal state distribution confirm thermalization of the molecules' rotational temperature in the nozzle. This is particularly evident from the temperature dependence of the \ket{5,5} state (see Fig.\;\ref{fig:popT}). Due to its high rotational energy of 241.2\,cm$^{-1}$, this state is less occupied at low temperatures.

It is desirable to characterize the internal state distribution with one basic quantity like a temperature. Finding a good description of the internal temperature of the guided beam is non-trivial. For a source temperature of 150\,K the mean rotational energy of the guided beam of 163.6\,cm$^{-1}$ would correspond to that of a thermal ensemble at a temperature of 155\,K. However, the mean rotational energy is not a good measure for the purity of the beam since by the nature of the filtering process, which depends on the Stark shift, the guided ensemble is non-thermal.

As a measure for the purity of the guided beam we hence determine the entropy $S=-\sum p_i\log p_i$ from the calculated populations $p_i$ of all guided states and compare this to the entropy of a thermal ensemble. In this calculation states up to $J$=12 were included, which was found to be sufficient. We define as entropic temperature $T_e$ of our guided beam the temperature at which a thermal gas has the same entropy. For a source temperature $T_s$ of 150\,K we find $T_e\approx$\,31\,K, for $T_s$=\,200\,K we find $T_e\approx$\,38\,K, and for $T_s$=\,300\,K we find $T_e\approx$\,47\,K. The purity can be even larger in the case of molecules with quadratic Stark shifts, such as D$_2$O, due to stronger selection in the filtering process \cite{RiegerD2O}.

In summary, we have shown that depletion spectroscopy is a powerful tool for measuring the internal state distribution of slow guided formaldehyde molecules. We find good agreement between the experimentally measured rotational state distribution and theoretical predictions based on source population and Stark shift calculations for varying source temperature. The method is not restricted to formaldehyde and should be widely applicable.

Depletion spectroscopy is not only a tool for internal state diagnostics. For example, it can be used for spectroscopy of species which are hard to study in the gas phase. Collinear spectroscopy in a cold guided beam has the advantage of long interaction times and good overlap with the slow molecular beam making it a good choice for studies of weak transitions. Since the electrostatic guide produces a continuous flux of molecules, it is a natural choice for combination with narrow-band cw lasers. Using a higher-order multipole guide, probing molecular transitions in more homogeneous electric fields is possible, resulting in narrow lines, and allowing for example measurements of dipole moments of electronically excited states. Furthermore, by applying a DC electric bias field, spectroscopy of oriented polar molecules should be possible.

The authors thank  T. Rieger for help in the initial stage of the experiment and M. Schmitt for assistance in fitting room-temperature spectra.
Support by EUROQUAM (Cavity-Mediated Molecular Cooling) and by the
Deutsche Forschungsgemeinschaft through the excellence cluster
"Munich Centre for Advanced Photonics" is acknowledged.

\end{document}